\documentclass[article,11pt,apl,amsmath,amssymb,showpacs,superscriptaddress]{revtex4-2}
\usepackage{epsf}      
\usepackage{graphicx}
\usepackage{color}
\usepackage{soul}
\usepackage{gensymb}
\usepackage{sidecap}
\usepackage{amsmath}
\usepackage{mathtools}
\usepackage{wrapfig}
\usepackage[hidelinks,colorlinks=true,linkcolor=blue,citecolor=blue]{hyperref}

\begin{document}
	\title{Unraveling the storage mechanism of Na$_{3}$V$_{2-x}$Ni$_x$(PO$_4$)$_2$F$_3$/C cathodes for sodium-ion batteries through electrochemical, {\it operando} X-ray diffraction and  microscopy studies}
	
	\author{Simranjot K. Sapra}
	\affiliation{Department of Physics, Indian Institute of Technology Delhi, Hauz Khas, New Delhi-110016, India}
	\affiliation{International College of Semiconductor Technology, National Yang Ming Chiao Tung University, 1001 University Road, Hsinchu 30010, Taiwan}
	\affiliation{Department of Materials Science and Engineering, National Yang Ming Chiao Tung University, 1001 University Road, Hsinchu 30010, Taiwan}
	
	\author{Jeng-Kuei Chang}
	\email{jkchang@nycu.edu.tw}
	\affiliation{International College of Semiconductor Technology, National Yang Ming Chiao Tung University, 1001 University Road, Hsinchu 30010, Taiwan}
	\affiliation{Department of Materials Science and Engineering, National Yang Ming Chiao Tung University, 1001 University Road, Hsinchu 30010, Taiwan}
	\affiliation{Department of Chemical Engineering, Chung Yuan Christian University, 200 Chung Pei Road, Taoyuan, 32023 Taiwan}
	
	\author{Rajendra S. Dhaka}
	\email{rsdhaka@physics.iitd.ac.in}
	\affiliation{Department of Physics, Indian Institute of Technology Delhi, Hauz Khas, New Delhi-110016, India}	
	\date{\today}

\begin{abstract}
The storage mechanism and diffusion kinetics of Na$_{3}$V$_{2-x}$Ni$_{x}$(PO$_{4}$)$_{2}$F$_{3}$/C ($x=$ 0--0.07) cathodes are investigated through electrochemical impedance spectroscopy (EIS), galvanostatic intermittent titration technique (GITT) and cyclic voltammetry (CV). All the samples are prepared through the facile pH-assisted sol-gel route and crystallize in the P4$_{2}$/mnm symmetry.  The optimal doping of Ni ($x=$ 0.05) exhibits superior specific capacities of 119 and 100 mAh g$^{-1}$ at 0.1 C and 10 C rates, respectively, along the excellent capacity retention of 78\% after 2000 cycles at 10 C rate with nearly 100\% Coulombic efficiency. The apparent diffusion coefficient values are found to be in the range of 10$^{-9}$--10$^{-10}$ cm$^{2}$ s$^{-1}$ through detailed analysis of CV and GITT. Moreover, we report the reversible structural evolution and morphological changes during charging and discharging under non-equilibrium conditions through the {\it operando} X-ray diffraction and the {\it in-situ} synchrotron based transmission X-ray microscopy, respectively. Further, to understand the stability mechanism and obtain precise polarization values, we performed the distribution of relaxation times (DRT) analysis using the EIS data. The structure and morphology are found to be stable after long cycling. 
\end{abstract}

\maketitle

\section{\noindent ~Introduction}
	
Sodium-ion batteries (SIBs) have rekindled research interest in the energy storage arena in view of ecological concerns and niche environmentally congenial alternatives, owing to their sustainable cost and abundant resources for large-scale stationary storage applications \cite{Larcher_NC_2015, Slater_AFM_2013, Yabuuchi_CR_2014, Eftekhari_JPS_2018}. However, finding suitable electrodes for high energy density as well as long cyclic life is a challenge due to the larger size of Na$^+$ (1.02 \AA) as compared to Li$^+$ (0.76 \AA). On the other hand, the SIBs also operate based on the intercalation chemistry similar to the commercial Lithium-ion batteries (LIBs), which, thus, instigates us to search for electrode materials analogous to LIBs \cite{Eftekhari_JPS_2018, Abraham_ELett_2020, Dwivedi_AEM_2021, Manish_CEJ_2023}. Also, for the battery operation, energy density, gravimetric capacity and cyclic/thermal stability are the critical factors, which rely on the choice of cathode material \cite{Hwang_CSR_2017, Sapra_Wiley_2021, Sapra_JMCA_2025}. A plethora of reported cathode materials primarily include layered metal oxides, organic compounds, honeycomb and polyanionic structures \cite{Hwang_CSR_2017, Pati_JPS_2024, Pati_JMCA_2022, Sun_AEM_2019, Manish_PRB_2024, Jian_AM_2017}. Among these, the polyanionic based Na Super Ionic Conductor (NASICON) materials, with general formula Na$_{x}$TM$_{2}$(XO$_{4}$)$_{3}$, TM = V, Fe, Cr, Mn, Ti etc. and X = P, S, Si, B, have become hotspot  due to their 3D open framework, which allows Na ions to be sodiated/desodiated under meagre lattice strains. Also, the high thermal and chemical stability, high operating potential and energy density due to the locked oxygen in PO$_{4}$$^{3-}$ unit and strong inductive effect, lend additional benefits to the overall electrochemical performance \cite{Sharma_CCR_25}. In this class, the Na$_{3}$V$_{2}$(PO$_{4}$)$_{2}$F$_{3}$ (NVPF)  have emerged as appealing cathodes for their improved and stable electrochemical performance \cite{Sapra_Wiley_2021, Masquelier_CRev_2013, Sapra_AMI_2024}. Interestingly, in the NVPF structure, the high Pauling's electronegativity of F$^{-}$ ion with high oxidative stability, coupled with the inductive effect of PO$_{4}$$^{3-}$ units, strengthens the ionicity of V--F bond and thereby, increase the working voltage \cite{Sharma_AEM_2020, Lin_CEJ_2023}. Notably, the NVPF exhibits a average working voltage of about 3.95 V and theoretical capacity of 128 mAh g$^{-1}$, corresponding to the reversible intercalation/deintercalation of 2 Na$^{+}$ ions, therefore, carries potential to possess energy density around 500 Wh Kg$^{-1}$ and serves as a key pillar in the family of cathodes \cite{Xu_MHoriz_2023, Gover_SSI_2006, Song_ECA_2014}. However, the concentration difference of F$^{-1}$ and PO$_{4}$$^{3-}$ groups vary the crystal structure and energy storage capabilities of the NVPF cathode \cite{Xu_MHoriz_2023, Li_ESM_2021}.
	
Nevertheless, the NVPF also inherits the defect of low intrinsic electronic conductivity and ionic diffusivity due to the insulating character of PO$_{4}$$^{3-}$ unit, which obstructs the electronic delocalization of --V--O--V-- \cite{Xu_MHoriz_2023}. This seriously affects the realization of optimal rate performance, which is primarily determined by the effective diffusion of Na$^{+}$ within the bulk material and across the electrode--electrolyte interface,  coupled with the efficiency of overall electronic conduction. In order to overcome with some of these issues, large number of strategic studies have been reported to tune the electronic conductivity and Na$^{+}$ diffusivity, for example, carbon coating, ion doping, electrolyte optimization and morphology control \cite{Xu_AEM_2018,Gu_ScienceB_2020, Dee_SCE_2021, Zhang_JMCA_2018, Dee_AMI_2024, Cai_JPS_2023, Guo_JEC_2022, Liu_JMCA_2017, Yi_NanoEnergy_2018, Zhou_Small_2024, Zhang_JPS_2024, Guo_ChemM_2025}, which substantially enhanced the rate and cycling performance. Nonetheless, the integration of cation doping with carbon coating is a mature strategy for enhancing rate performance, attributable to the synergistic interplay of induced vacancies, lowered ion diffusion barriers, and increased intrinsic electronic conductivity \cite{Wang_JMCA_2024, Xiao_SMALL_2023}. To understand the doping effects, a few recent research reports have focused on the positive aspects of Ni$^{2+}$ in various polyanionic frameworks. For example, the Ni$^{2+}$ substitution in LiFePO$_{4}$/C composites increased the crystallinity, retaining the olivine phase and showed capacity retention of 93.8\% after 200 cycles at 10 C rate \cite{Liu_CI_2020}. Li {\it et al.} also designed the Ni$^{2+}$ substituted NVP series and found significant enhancement in the capacity upon doping, owing to increased Na$^{+}$ ion concentration with negligible structure change. It exhibited 95.5\% capacity retention after 100 cycles in Na$_{3.03}$V$_{1.97}$Ni$_{0.03}$(PO$_{4}$)$_{3}$/C sample \cite{Li_AMI_2016}. Bag {\it et al.} compared the Mg$^{2+}$, Ni$^{2+}$ and Co$^{2+}$ dopants in NVP crystal structure and discussed in detail the impacts of doping for the improved electrochemical performance \cite{Bag_AEM_2020}. Recently, the impact of Ni$^{2+}$ doping in NVP is studied in hybrid lithium-ion battery systems to understand the synergic effect on transport of Na$^{+}$ and Li$^{+}$ \cite{Akhtar_AEM_2021}. The optimal doping delivered an impressive specific capacity of 230 mA g$^{-1}$ against Li$^{+}$ and 106 mAh g$^{-1}$ against Na$^{+}$ \cite{Akhtar_AEM_2021}. Interestingly, for the LIBs, the Ni$^{2+}$ is already realized to be a crucial element especially LiNi$_{0.8}$Co$_{0.1}$Mn$_{0.1}$O$_{2}$ (NMC811), LiNi$_{0.5}$Mn$_{1.5}$O$_{4}$ (LNMO) owing to its ability to undergo multiple redox reactions, thermal stability, high cell potential, cost-effectiveness, and versatility to be incorporated in different structural frameworks \cite{Manthiram_Nature_Com_2020}. However, to the best of our knowledge, detailed investigation to understand the effect of Ni$^{2+}$ dopant in NVPF cathode is largely lacking. 

Therefore, in this work, taking into account some of the positive attributes, bivalent Ni$^{2+}$ ions are incorporated at the V$^{3+}$ site along with the carbon coating to yield Na$_{3}$V$_{2-x}$Ni$_{x}$(PO$_{4}$)$_{2}$F$_{3}$/C ($x=$ 0--0.07) samples. Also, to circumvent the volatile nature of fluorine and its loss during synthesis, the pH of the precursor solution during the synthesis was controlled between 4.2 and 8.0 to prevent hydrogen fluoride (HF) formation in acidic media \cite{Deng_NanoEnergy_2021, LongLi_JMCA_2022}. Notably, we report in-depth investigation of the diffusion kinetics and electrochemical performance using cyclic voltammetry (CV), galvanostatic cycling, electrochemical impedance spectroscopy (EIS), and galvanostatic intermittent titration technique (GITT). More importantly, variations in crystal structure and morphology during charging-discharging process are studied through {\it operando} X-ray diffraction and {\it in-situ} synchrotron based X-ray microscopy techniques. 

\section{\noindent ~Experimental Section}

\noindent 2.1 \textit{~Material synthesis:}\\
The Na$_{3}$V$_{2-x}$Ni$_{x}$(PO$_{4}$)$_{2}$F$_{3}$/C ($x=$ 0--0.07) samples are prepared by the conventional sol-gel method, as illustrated in Fig.S1 of \cite{SI}. The precursors, i.e., sodium fluoride (NaF, Acros Organics, $\mathrm{>}$99\%), ammonium dihydrogen phosphate (NH$_{4}$H$_{2}$PO$_{4}$, Showa Chemicals, $\mathrm{>}$99\%), nickel nitrate hexahydrate (Ni(NO$_{3}$)$_{2}$.6H$_{2}$O, Showa Chemicals, $\mathrm{>}$99.9\%) and ammonium metavanadate (NH$_{4}$VO$_{3}$, Strem Chemicals, $\mathrm{>}$99\%) were used accordingly to the stoichiometric ratio with 10 wt.\% excess NaF to compensate for Na and F loss during heat treatment. The citric acid was chosen as the reducing and conducting agent for reducing V$^{+5}$ to V$^{+3}$ valence state and aids in the carbon coating, respectively. The stoichiometric amounts of the above precursors were dissolved in deionized water and stirred overnight at room temperature until the solution turned dark-green. The pH of the solution was controlled at a value of 8.0 to maintain the basicity of the solution, which prevents fluorine loss due to HF formation during the gelation. The resultant solution was allowed to evaporate at 70\degree C, and the formed gel was dried overnight in an oven at 120\degree C. The dried powder was calcined in a tube furnace at 350\degree C for 4 hrs and 600\degree C for 8 hrs in an argon atmosphere, followed by washing in deionized water and drying, to yield the fluorophosphates. 
	
\noindent 2.2 \textit{Electrode and coin-cell fabrication:}\\
A cathode slurry was prepared by mixing 80 wt.\% active material, 10 wt.\% Super P (Timcal), and 10 wt.\% polyvinylidene fluoride(PVDF, Sigma-Aldrich) in N-methyl-2-pyrrolidone solution (NMP, Sigma Aldrich). This slurry was pasted onto Al foil (current collector), vacuum-dried at 90\degree C for 10 hrs for electrode fabrication. The electrode was punched for the CR2032 dimensions, and an active mass loading of $\mathrm{\sim}$2.5--3.0 mg cm$^{-2}$ was obtained. The half cells were assembled using a thin Na metal foil (Sigma-Aldrich) as counter electrode, glass fiber membrance (Advantec, Japan) as the separator and 1 M NaClO$_{4}$ in ethylene carbonate (EC, Kishida Chemicals) and polypropylene carbonate (PC, Kishida Chemicals) with the 10 wt.\% fluoroethylene carbonate (FEC, Sigma Aldrich) additive as the electrolyte in an argon-filled glove box (Vitro, Innovation Technology Co. Ltd.), where both the moisture and oxygen content levels were maintained at $\mathrm{\sim}$0.3 ppm. 
	
\noindent 2.3. \textit{Materials characterization:}\\
The high resolution synchrotron X-ray diffraction (XRD) data are collected with $\lambda=$ 0.61992 \AA~ using a Mythen detector, determined by the LaB$_{6}$ standard at TPS 19A, National Synchrotron Radiation Research Center (NSRRC), Hsinchu, Taiwan. The powder XRD measurements are conducted with Cu K$_{\alpha}$ radiation (wavelength of 1.5418~{\AA}) using Bruker D2 Phaser and Panalytical Xpert$^{3}$ diffractometers. The sample morphology was examined with the field emission scanning electron microscope JEOL-FESEM (EVO18 and JSM7800F Prime), and transmission electron microscopy (TEM, JEOL F2100F). The energy-dispersive X-ray spectroscopy (EDS) was employed to probe the chemical composition. The Raman modes were studied with the micro Raman spectrometer (LabRAM HR 800) with a probing wavelength of 514 nm. The vibrational modes were measured with the Fourier Transmission Infrared Spectroscopy (FTIR, Thermo Nicolet-IS-50) in the spectral range of 500--2000 cm$^{-1}$ by preparing a 13 mm pellet of the powder sample with the potassium bromide, KBr.  The X-ray photoelectron spectroscopy, XPS (Thermo VG-Scientific ESCALAB Xi$^{+}$ with Al K$_{\alpha}$ source h$\nu=$ 1486.6 eV) was used to examine the surface electronic structure. The core level binding energy values are corrected by fixing C 1$s$ at 284.6 eV and analyzed in IGOR Pro software using Lorentzian and Gaussian line profiles. 

\noindent 2.4. \textit{Electrochemical measurements:}\\
The charge and discharge performances of the fabricated cells were evaluated between 2.0 and 4.3 V with various currents using Neware BTS8.0 battery tester at room temperature. The electrochemical impedance spectroscopy (EIS) measurements were conducted in the 100 kHz to 10 mHz frequency range with an AC potential perturbation amplitude of 10 mV using VSP300 potentiostat The cyclic voltammetry (CV) at a sweep rate of 0.1--1.0 mV s$^{-1}$ and {\it in-situ} EIS data with cycling in the frequency range of 10 kHz--10 mHz were collected using the BT-LAB V1.64 potentiostat. The Galvanostatic intermittent titration technique (GITT) measurements were conducted using the Arbin battery cycler (BT-2043) with a pulse time of 20 mins and a relaxation of 3 hrs at a 0.1 C rate. The {\it operando} X-ray diffraction measurements are performed in an electrochemical cell with the Be window and is subjected to X-ray radiation during charging/discharging at a 0.3 C rate using a Bruker D2 phaser and Biologic SP-150 potentiostat at room temperature. The {\it in-situ} transmission X-ray microscopy (TXM) measurements were performed at TLS BL01B1 Beamline, NSRRC Hsinchu, Taiwan. In order to have better resolution and X-ray penetration, the free standing electrode was fabricated using 70 wt.\% of the PVDF binder, 20 wt.\% of active material, and 10 wt.\% of carbon additive (Super-P). The galvanostatic charging and discharging was conducted using SP-150 potentiostat where the TXM images were recorded after every 10s. The TXM image is formed when X-rays pass through the zone plate optical system and a phase ring. The distribution of relaxation times (DRT) analysis for the impedance data at various cycles was performed using an open source GUI Interface (DRT tools, MATLAB) \cite{Wan_ECA_2015}. 		
\begin{figure*}
\includegraphics[width=\linewidth]{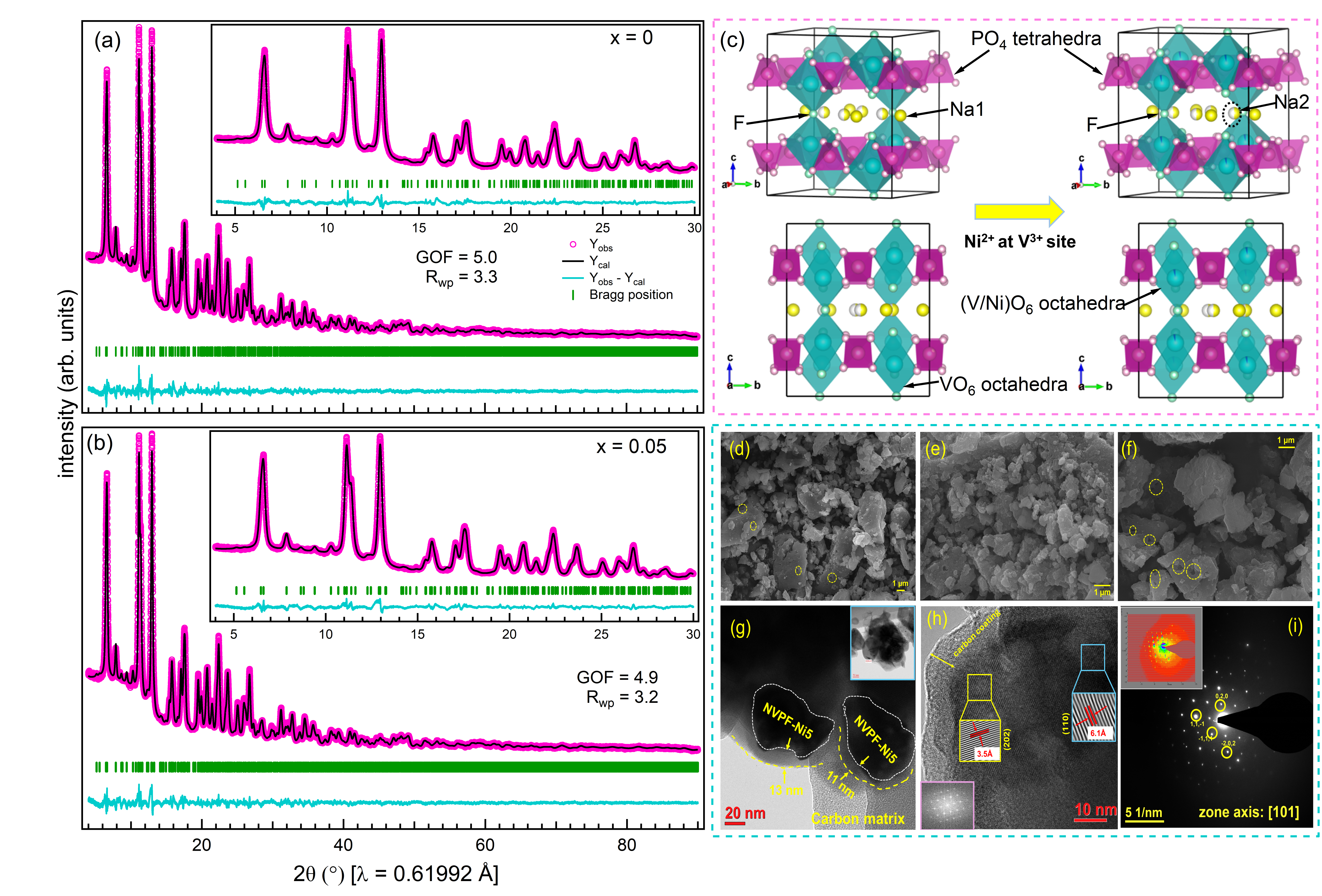}
\caption{(a, b) The synchrotron X-ray diffraction patterns of the Na$_{3}$V$_{2-x}$Ni$_{x}$(PO$_{4}$)$_{2}$F$_{3}$/C ($x=$ 0, 0.05)/C powders with rietveld refined profiles; (c) the crystal structure of the NVPF with the Ni$^{2+}$ substituted at the V$^{3+}$ site, where the Na1, Na2, V, Ni, P, O, F are shown by the yellow, mixed yellow-white, teal, blue, dark purple, light pink and light green colors; the high resolution FE-SEM image for the (d) $x=$ 0 and (e) $x=$ 0.05 with its zoomed view at 10 KX in (f); the high resolution TEM images for the $x=$ 0.05, showing particles and carbon coating (cuboid-like particle in inset) in (g), and corresponding diffraction gratings with $d$--spacing calculations and FFT (inset) in (h) and the SAED pattern (i) with the corresponding surface profile in inset.}
\label{xrd}
\end{figure*}

\section{\noindent ~Results and discussion}
	
\noindent 3.1 \textit{Crystal and electronic structures, morphology and microstructure with elemental mapping:} \\ 
The high resolution synchrotron X--ray diffraction patterns and the corresponding Rietveld refined profiles for the $x=$ 0 and 0.05 samples are depicted in Figs.~\ref{xrd}(a, b), which clearly show sharp peaks and proves the highly crystalline structure. We find slight increase in the lattice parameters with $x$ due to larger ionic radius of Ni$^{2+}$ (0.69 \AA) as compared to  the V$^{3+}$ (0.64 \AA) \cite{Shannon_Crystal_1976} which enhance the interstitial space and facilitate the faster Na$^{+}$ diffusion. The goodness of fit (GOF) and R$_{w}$ values are also mentioned in Figs.~\ref{xrd}(a, b). Moreover, the detailed structural comparison for all the samples, i.e., Na$_{3}$V$_{2-x}$Ni$_{x}$(PO$_{4}$)$_{2}$F$_{3}$/C ($x=$ 0--0.07), is presented in Figs.~S2(a--e) of \cite{SI} along with the Rietveld refinement of laboratory based XRD patterns. The reliability factors, R$_{wp}$ and R$_{p}$ and goodness of fit, $\chi$$^{2}$ derived from the refinement, are about 11, 12 and 2\% respectively, hence proving the accuracy of the refinement procedures. The retrieved crystallographic parameters and wyckoff positions from the refinement are summerized in Table~S1 and Tables S2--S6 of \cite{SI}, respectively. Also, the Na--O bond length is found to increase with $x$, whereas the P--O and V--O bond lengths are reduced, thereby creating open framework with larger diffusion channels and stabilizing tetragonal crystal symmetry (space group: P4$_{2}$/mnm) \cite{Meins_SSC_1999}. There are no impurity peaks observed and parent crystal structure is maintained for all the samples. Interestingly, it is observed that splitting between (111) and (002) peak is weak, in comparison to the standard NVPF sample, which signifies the shortened growth along (111) peaks arises from the precursor, NaF. It acts as crystal growth inducer and catalyses the reaction towards certain crystallographic directions and therefore, affects morphology and crystal structure \cite{Zhu_JMCA_2020, Guo_JEC_2022, Zhang_JPS_2024}. 

The calculated values of deviation for Na and V using mathematical model equations S1 and S2 are found to be 1.98 and 0.25, respectively, which implies Ni$^{2+}$ ion prefers to occupy the V$^{3+}$ site instead of Na$^{+}$ site. All the samples represent a pseudo-layered structure in which the Na and F atoms form one Na-F layer, while VO$_{4}$F$_{2}$ octahedra and PO$_{4}$ tetrahedra form the second layer and both these layers are repeated in an alternate manner along the {\it c-} axis and constructs the crystal framework \cite{Park_AFM_2014, Meins_SSC_1999}. In this repetitive two layer slabs, two VO$_{4}$F$_{2}$ octahedra connect through the F atom to form V$_{2}$O$_{8}$F$_{2}$ bi-octahedra, which are connected to PO$_{4}$ tetrahedra via O atom in {\it a-b} plane. This 3D open framework provides large and broad channels for Na ions along the {\it a} and {\it b} directions, facilitating the smooth ionic transport, as shown in Fig.~\ref{xrd}(c). Here, the Na ions occupy two different coordination sites, Na(1) and Na(2). The Na(1) site is surrounded by two F atoms and four O atoms, known as triangular prismatic site while Na(2) site, denoted as augmented triangular site, is attached to the F atom at the apex-square pyramid. Na$^{+}$ ions are located at two distinct Wyckoff positions (8h and 8i) and occupancy of the Na(1) and Na(2) in the crystal structure is 1 and 0.5. 

Figs.~\ref{xrd}(d--f) show the FE-SEM images of $x=$ 0 and 0.05 samples where the morphology indicates the collection of ill-defined primary particles fused together. For the $x=$ 0.05 sample we clearly see in Fig.~\ref{xrd}(f) that the particle sizes are heterogeneously distributed in the range of 1--5 $\mu$m, which can alters the effective ion transport during battery cycling \cite{Guo_ChemM_2025}. The FE-SEM images of other samples are presented in Figs.~S3(e--g) of \cite{SI}, which also confirm the agglomeration and anisotropy leading to the pseudo-cuboid like irregular morphology. 
The irregular morphology of the final product also arises from the complexation reactions of citric acid with the vanadium during synthesis and addition of ammonium hydroxide which modifies the surface properties of the particles during the nucleation and growth \cite{Zhang_JMCA_2018}. The pH modification strategy creates a porous structure in the crystal, as evidenced by the highlighted circles in Figs.~\ref{xrd}(d, f), which facilitates enhanced interfacial ionic flux between the electrode and electrolyte. Speculatively, these mesopores can provide more active sites and further cushion the volume changes during the Na$^{+}$ intercalation/deintercalation. Yi {\it et al.} reported that surface and adsorption energies mainly control the morphology during sample synthesis \cite{Yi_ELett_2019}. Also, Zhu {\it et al.} observed that NaF acts as a growth inducer and orients the particles along certain facets, which result in the pseudo-cuboid like morphology \cite{Zhu_MCF_2020}.  Also, adsorption of NaF lowers the surface energy and stabilises the [001] crystal facet, therefore orienting the growth of NVPF crystals along [001] direction \cite{Yang_Nature_2008}. This effect is further controlled by the alkaline environment (pH=8), where adsorption of OH$^{-}$ ions lowers the surface energy of all the crystal facets, thereby yielding the irregular dimensional particles. 

\begin{figure*}
 \includegraphics[width=\linewidth]{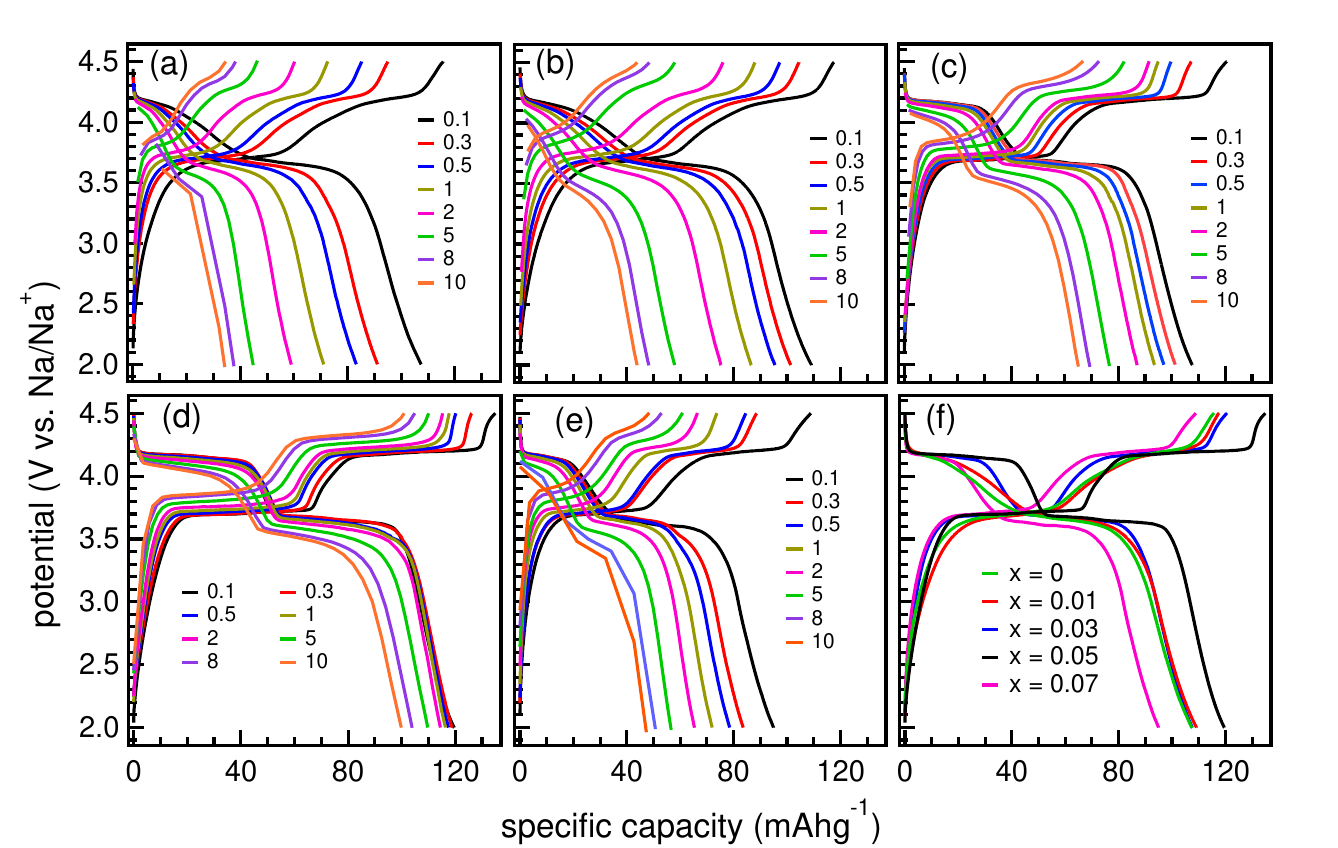}
\caption{(a-e) The galvanostatic charge-discharge (GCD) characteristics at the different C-rates (0.1--10 C) for the Na$_{3}$V$_{2-x}$Ni$_{x}$(PO$_{4}$)$_{2}$F$_{3}$/C ($x=$ 0--0.07) cathodes. (f) the comparison of GCD profiles measured at 0.1 C. } 
\label{GCD}
\end{figure*}

Furthermore, for the $x=$ 0.05 sample the high resolution TEM images are shown in Figs.~\ref{xrd}(g, h), which indicate that the particles are distributed in the carbon matrix and surrounded by the amorphous carbon coating of thickness 11--13 $\mu$m on their surface. The gas evolution from citric acid decomposition during synthesis promotes the formation of a porous carbon layer on the active particle surfaces. This is expected to enhance both electrolyte penetration and surface conductivity, enabling the movement of the Na$^+$ and electrons and ultimately leading to improved rate performance \cite{Gu_ScienceB_2020}. The data in Fig.~\ref{xrd}(h) shows the distinct and vivid lattice fringes and carbon coating with $d-$spacing of 3.5 \AA~and 6.1 \AA~corresponding to (202) and (110) crystal planes of NVPF, respectively. The inset in Fig.~\ref{xrd}(h) shows the Fourier transform of the region, marked by yellow plane (202), which confirms the crystalline structure. The selected area electron diffraction (SAED) pattern, in Fig.~\ref{xrd}(i) indicates good contrast with the bright spots, corresponding to the lattice planes (020), (11-1), (-111), and (-202) along the zone axis [101]. The inset shows the surface profile of the SAED pattern where each bright spot is corresponding to the lattice plane and therefore, confirms the crystallinity of $x=$ 0.05 sample. Further, the elemental mappings are done through high-angle annular dark-field scanning transmission electron microscopy (HAADF--STEM), see Fig.~S3(a) of \cite{SI}. This confirms that elements Na, V, P, F, O and Ni are uniformly distributed across the selected region and Ni has been successfully incorporated into the crystal lattice. Also, further characterizations using  FTIR, Raman and XPS are presented in Figs.~S4 of \cite{SI}, which confirm the structural coordination, presence of coated carbon and the stretching/bending modes as well as surface chemical states of the elements present. 

\begin{figure*}
	\includegraphics[width=0.95\linewidth]{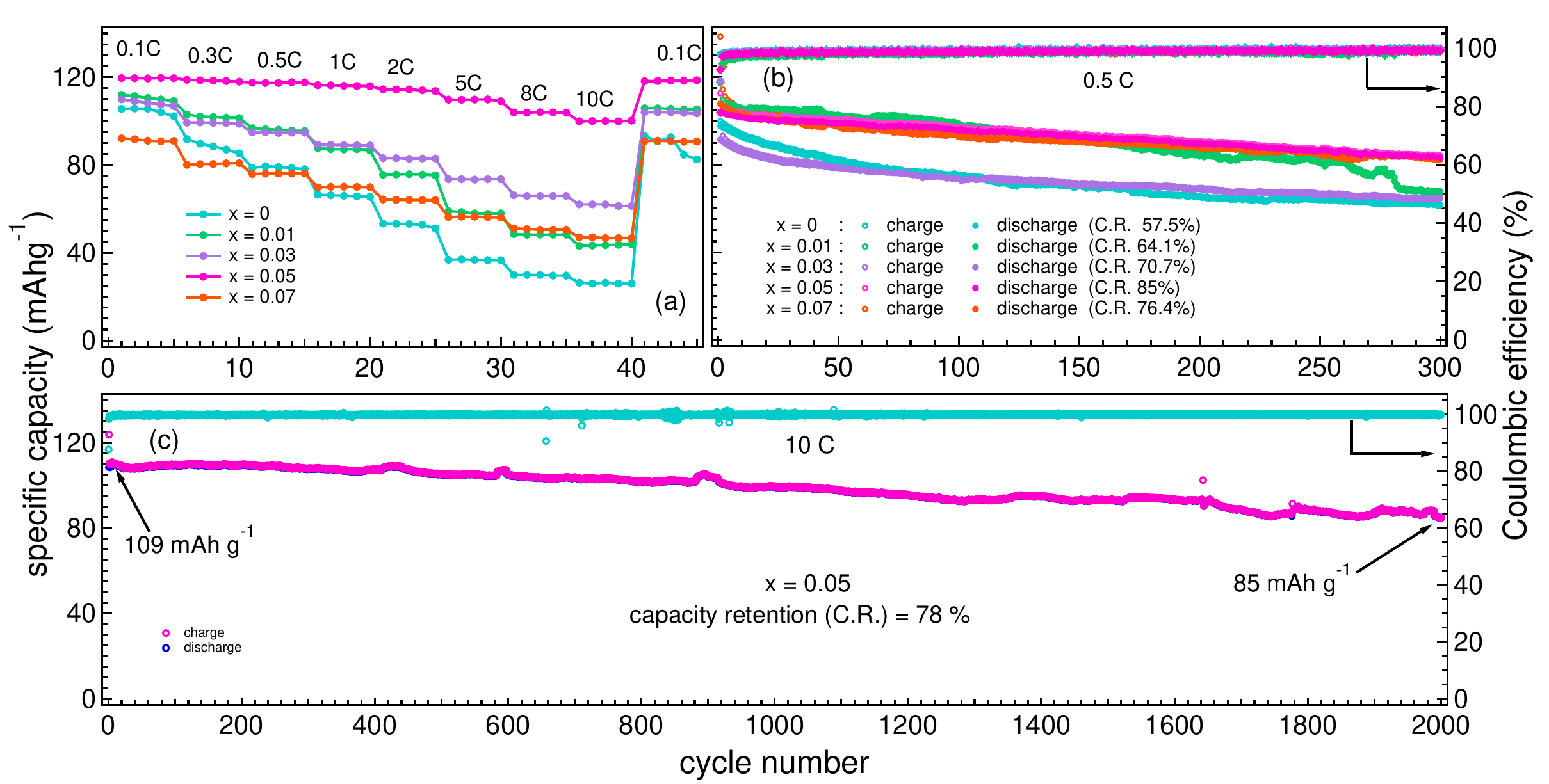}
	\caption{(a) The rate performance at different C rates, the long cyclic performance (b) at 0.5 C rate up to 300 cycles for the Na$_{3}$V$_{2-x}$Ni$_{x}$(PO$_{4}$)$_{2}$F$_{3}$/C ($x=$ 0--0.07) cathodes, and (c) at 10 C rate up to 2000 cycles for the $x=$ 0.05.}
	\label{LC}
\end{figure*} 

\noindent 3.2 \textit{Electrochemical Performance and diffusion kinetics:}\\ 
The galvanostatic charge-discharge (GCD) profiles are measured in the voltage window of 2.0--4.5 V at different current rates between 0.1 and 10 C, depicted in Figs.~\ref{GCD}(a--e). The GCD curves exhibit two symmetric potential plateaus at 3.7 V and 4.2 V, yielding an average operating potential of 3.95 V, which is higher than its sister compound, NVP (3.4 V), owing to its stronger inductive effect from F$^{-}$ ions. The potential plateau at 3.7 V corresponds to the extraction of Na$^{+}$ from the Na(2) site and the plateau at 4.2 V corresponds to the extraction of Na$^{+}$ from the Na(1) site, with the extraction of two Na$^{+}$ per formula unit \cite{Song_ECA_2014}. Notably, by substituting the Ni$^{2+}$ ions at the V site, the 4.2 V voltage plateau got stabilized with lowering the polarization and the specific capacity improved from 107 mAh g$^{-1}$ at 0.1 C for pristine cathode to around 119 mAh g$^{-1}$ for the $x=$ 0.05 concentration, which found to be much closer to the theoretical values of around 128 mAh g$^{-1}$. The GCD profiles measured at 0.1 C are compared in Fig.~\ref{GCD}(f). More importantly, a remarkable enhancement is observed at 10 C with the specific capacity value of 100 mAh g$^{-1}$ for the $x=$ 0.05 cathode, which may arise from the partial oxidation of V$^{4+}$ to V$^{5+}$ \cite{Lin_CEC_2022}. Nonetheless, the initial capacity loss may be credited to the electrolyte decomposition on the electrode surface at higher potentials, formation of cathode electrolyte interface (CEI) and solid electrolyte interface (SEI) and other parasitic reactions which happen during first few cycles \cite{Verma_ECA_2010, Guo_ChemM_2025, Zhang_JPS_2024}. Intriguingly, the excellent rate performance of all the Ni doped cathodes at the different current rates (0.1--10 C), shown in Fig.~\ref{LC}(a). It clearly indicates a remarkable increase in the specific capacity at higher rates with the increase in Ni$^{2+}$ doping particularly for the $x=$ 0.05 concentration. This is fundamentally governed by the Na$^{+}$ diffusivity and electronic conductivity [see Fig.~S5(f) of \cite{SI}] where the diffusion of Na$^{+}$ is affected by bulk resistance as well as the charge transfer resistance across the cathode-electrolyte interface and through the liquid electrolyte. Further, after cycling the cells at high rates up to 10 C, almost initial capacity can be achieved back at 0.1 C,  demonstrating the structural reversibility of all the cathodes, as seen in Fig.~\ref{LC}(a). Interestingly, when doping concentration exceeds the value of 0.05, the specific capacity begins to decline as a result of decrease in the electronic conductivity, see Fig.~S5(f) of \cite{SI}, showing the volcanic shaped trend. 

In order to test the battery life, Fig.~\ref{LC}(b) shows the cyclic performance measured at 0.5 C where we find that the $x=$ 0.05 electrode retained around 85\% specific capacity with nearly 100\% Coulombic efficiency for 300 cycles. Conversely, the capacity retention of 57.5\%, 64\%, 71\% and 76\% is observed for the $x=$ 0, 0.01, 0.03 and 0.07 electrodes, as shown in Fig.~\ref{LC}(b). Intriguingly, the stability tests of $x=$ 0.05 electrode measured at 10 C is presented in Fig.~\ref{LC}(c), which exhibit excellent initial capacity of around 109 mAh g$^{-1}$ and retention of 78\% maintaining the capacity of around 85 mAh g$^{-1}$ at the end of 2000 cycles with an average Coulombic efficiency of $>$99.8\%. Overall, it has been observed that Ni doped electrodes exhibit better cycling performance than the pristine, which can be credited to the increased conductivity and structural stability upon Ni$^{2+}$ incorporation into lattice. The shortened V--O and P--O bond lengths correspond to the stronger covalent bonding, which further aids in stabilizing the lattice oxygen and hence, enhanced structural stability \cite{Dee_AMI_2024, Guo_JEC_2022}. 

\begin{figure*}
\includegraphics[width=\linewidth]{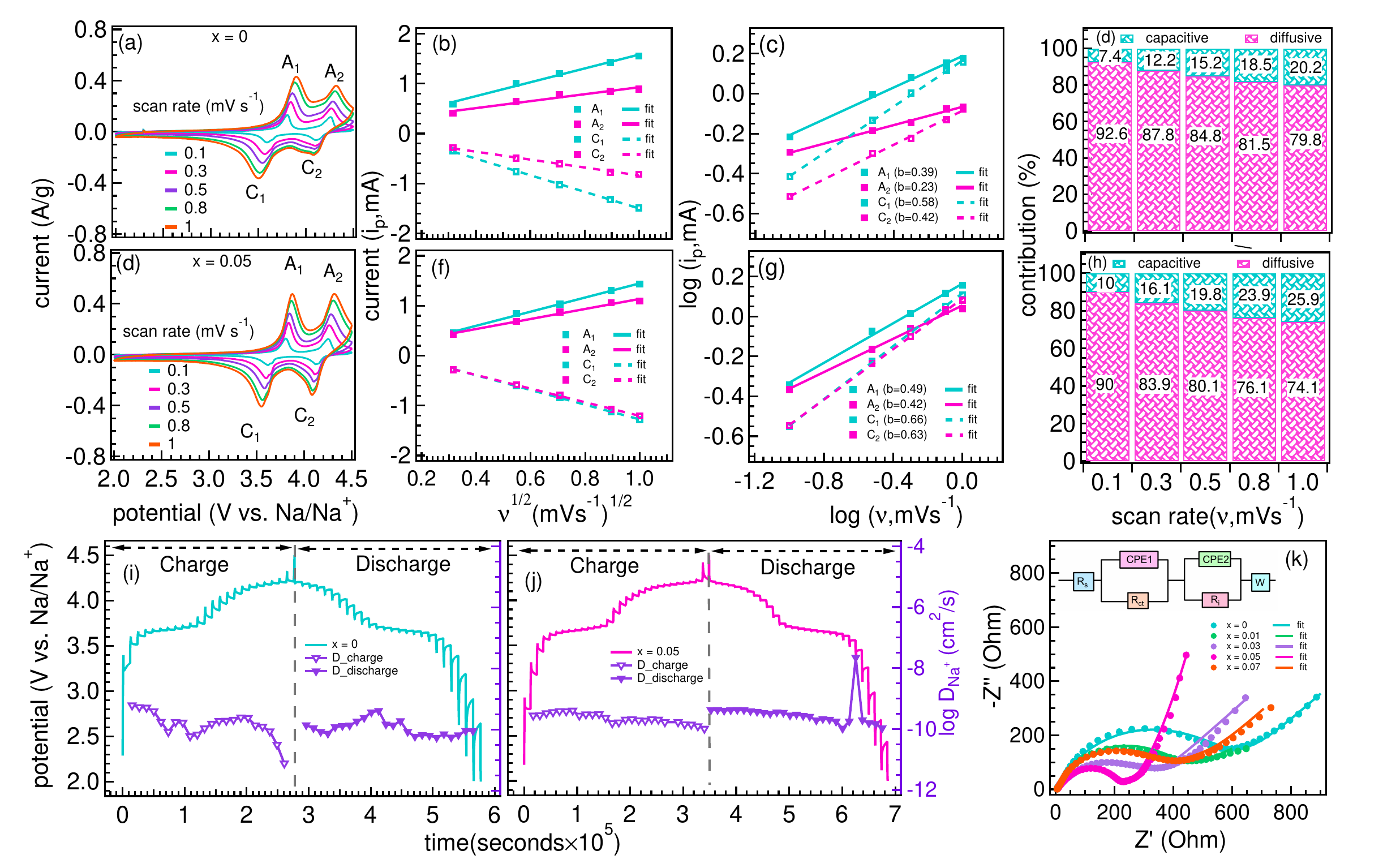}
\caption{The cyclic voltammogram (CV) at scan rates from 0.1--1.0 mVs$^{-1}$, the peak current versus scan rate, the log of peak current versus log of scan rates, and the contribution of the capacitive and diffusive components from the total current at different scan rates, for the pristine (a--d) and $x=$ 0.05 (e--h) cathodes, respectively; the GITT profiles and the diffusion coefficients during the charging and discharging measured at 0.1 C rate in the 2.0--4.5 V range for the (i) pristine and (j) $x=$ 0.05 cathodes, (k) the Nyquist Plot (Re Z vs -Imz Z) for all the cathodes ($x=$ 0--0.07) with Randles circuit in the inset.} 
\label{CV}
\end{figure*}
 
To elucidate the effect of Ni$^{2+}$ substitution on the reaction kinetics, the cyclic voltammetry (CV) measurements are performed in a potential window of 2.0--4.5 V, ranging from a scan rate of 0.1 mV s$^{-1}$ to 1.0 mV s$^{-1}$, as depicted in Figs.~\ref{CV}(a, d) for the $x=$ 0 and 0.05 cathodes, respectively, and others are shown in Fig.~S5 of \cite{SI}. Herein, we observe the anodic (A$_{1}$, A$_{2}$) and cathodic (C$_{1}$, C$_{2}$) peaks, corresponding to the reversible intercalation/deintercalation of Na$^{+}$ through V$^{3+}$/V$^{4+}$ redox couple, which found to be consistent with plateaus in the charge-discharge curves, shown in Figs.~\ref{GCD}(a--e). We observe a slight shift in opposite direction for the oxidation and reduction peaks with increasing the scan rate, which signifies an increase in polarization \cite{Sapra_AMI_2024}. However, the polarization value for the $x=$ 0.05 cathode is found to be less, see Table S7 of \cite{SI}, which is consistent with its significantly improved electrochemical performance as compared to the others. Furthermore, the current values of oxidation/reduction peaks increases with the scan rates and follow the linear relation as a function of $\nu$$^{1/2}$, depicted as Randles-Sevcik plot in Fig.~\ref{CV}(b, f) respectively. Assuming that current obeys the power law relation, i$_p$ = $a$$\nu$$^{b}$; where $a$ and $b$ are constants, i$_p$ and $\nu$ are the peak current and the scan rate. On taking the logarithm, the value of $b$ (indicator of type of Na$^{+}$ (de)intercalaton mechanism) is calculated from the slope of the log($i_p$) and log($\nu$) graph, as depicted in Figs.~\ref{CV}(c, g). As reported in the literature \cite{Mathis_AEM_2019, Babu_AEM_2020, Liu_AdvSci_2018}, if the value of $b$$\leq$ 0.5, the charge storage mechanism is diffusion-controlled (Faradaic) and the value of $b=$ 1.0 corresponds to the surface-controlled (capacitive) charge storage mechanism. If the $b$ value lies between 0.5 and 1.0, ion storage is termed as pseudo--capacitive (surface + diffusion). 

Herein, the values of $b$ are 0.39 and 0.23 for the anodic peaks (A$_{1}$, A$_{2}$), and 0.58 and 0.43 for the cathodic peaks (C$_{1}$, C$_{2}$) of the pristine cathode; while 0.49 and 0.42, and 0.66 and 0.63 for the $x=$ 0.05 signifying semi-infinite linear diffusion, which enables the fast electrochemical reaction dynamics \cite{Liu_AdvSci_2018, Mathis_AEM_2019, Babu_AEM_2020}. The contributions can be quantified by the sum of capacitive (k$_{1}$$\nu$) and diffusion controlled (k$_{2}$$\nu$$^{1/2}$) through the equation i(V) = k$_{1}$$\nu$ + k$_{2}$$\nu$$^{1/2}$ and the values are depicted in Figs.~\ref{CV}(d, h), which are consistent with the $b-$values and signify the contributions from diffusion dynamics and surface controlled behavior. In addition, the apparent diffusion coefficient of sodium ions can be determined through the Randles-Sevcik equation (\ref{RS_CV}), as below \cite{Yi_NanoEnergy_2018}:
\begin{equation}
i_{p} = 2.69\times 10^{5}n^{3/2}ACD^{1/2}\nu^{1/2}
\label{RS_CV}
\end{equation}
where $i_p$ is the peak current (in mA), $n$ is the number of electrons transferred in the redox reaction ($n=$ 2), $A$ is the area of the electrode (1.33 cm$^{2}$), $C$ is the bulk concentration of the electrode (in mol cm$^{-3}$), $D$ is the diffusion coefficient of the Na ions (in cm$^{2}$ s$^{-1}$) and $\nu$ is the scan rate (in mV s$^{-1}$). The calculated values of diffusion coefficient, D$_{Na^{+}}$ are found to be in the range of 10$^{-9}$--10$^{-10}$ cm$^{2}$ s$^{-1}$ for the anodic and cathodic peaks. 

It is important to note that the mobility of Na$^+$ inside the electrode is a critical parameter, which affects the electrochemical performance and cycle life of the SIBs. Therefore, GITT experiments were conducted in the potential window of 2.0--4.5 V by applying current pulse ($\tau$) of 20 min and relaxation time of 180 min for the uniform distribution of Na$^+$ to reach electrochemical quasi-equilibrium OCV at 0.1 C rate. During the measurement, the potential of the electrode varies upon application of the current due to the ohmic polarization, which is a transient process and when the current is constant, the potential decreases gradually, due to which the transient voltage, dE$_{\tau}$ does not include the voltage contribution from the ohmic drop. The experiment is repeated until fully discharged voltages are reached and from each ``pulse--relaxation'' unit, the state-of-charge (SOC) diffusion coefficients is derived. The galvanostatic titration charge--discharge curves are depicted in Figs.~\ref{CV}(i, j) for the $x=$ 0 and 0.05 cathodes, respectively. The apparent sodium diffusion coefficient (D$_{Na^{+}}$) values are calculated according to Fick's second law of diffusion \cite{Rui_ECA_2010, Morais_MaterAdv_2024}:
\begin{equation}
D_{Na^{+}} =\frac{4}{\pi \tau }(\frac{m_{B}V_{M}}{M_{B}A})^2(\frac{{\Delta E}_S}{\tau \left(\frac{dE_{\tau }}{d\sqrt{\tau }}\right)})^2,\ \ \tau \ll L^2/D_{{Na}^+}  
\end{equation}
Considering the linear variation of the transient voltage (E$_{\tau}$) with $\tau^{1/2}$; the above equation can be written as:
\begin{equation}
	D_{Na^{+}} =\frac{4}{\pi \tau }(\frac{m_{B}V_M}{M_{B\ }A})^2(\frac{{\Delta E}_S}{\Delta E_{\tau }})^2 
\end{equation} 
where the m$_{B}$, V$_{M}$ and M$_{B}$ are the active mass loading, molar volume and molecular weight of the active material of the cathode, L is the thickness and A is the active surface area of the electrode, $\tau$ is the pulse time in which constant current is applied, $\Delta $E$_{S}$ and $\Delta $E$_{\mathrm{\tau}}$ represent the difference before and after the current pulse and potential difference between the equilibrium potential and the potential maximum at the end of current pulse. The calculated average values of D$_{Na^{+}}$ for the charging and discharging process are in the range of 10$^{-9}$--10$^{-10}$ cm$^{2}$ s$^{-1}$ with somewhat better for the Ni$^{2+}$ substituted electrode, which found to be consistent with observation of improved specific capacity of $x=$ 0.05 concentraion. 

\begin{figure*}
	\includegraphics[width=\linewidth]{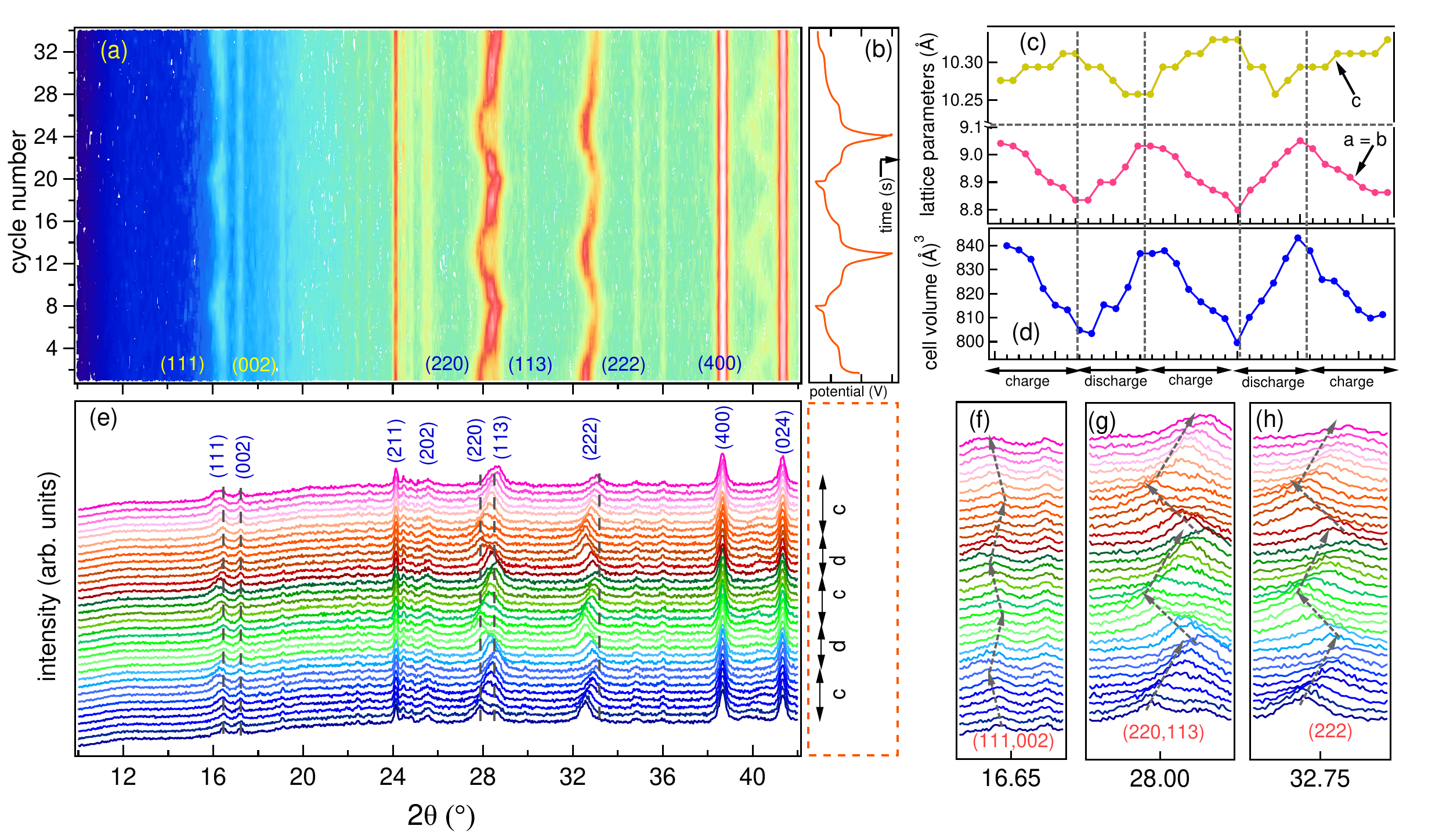}
	\caption{(a) The contour 2D plot of the {\it operando} XRD experiment, (b) the galvanostatic charge and discharge curves for first cycle charge--discharge, second cycle charge--discharge and third cycle charge, the variation of (c) the lattice parameters ($a$ and $c$) and (d) the cell volume (V) during the sodiation and desodiation process; (e) the recorded XRD patterns on the $x=$ 0.05 cathode while charging (denoted as c) and discharging (denoted as d) at 0.3 C rate with the step size of 0.03$\degree$ alongwith the zoomed view of (111, 220, 222) diffraction peaks in (f--h). }
	\label{operando}
\end{figure*}

 Moreover, the electrochemical impedance spectroscopy (EIS) measurements are performed to probe the electrode kinetics across the electrode-electrolyte interface. The Nyquist plot is presented in Fig.~\ref{CV}(k), where the inset shows the analogous Randles circuit, which uncover the resistances involved in the migration of the Na ions. The intercept of the semicircle signifies the solution and contact resistance (R$_{s}$) at high frequencies and a semicircle is described by the parallel RCPE (resistor$\parallel$constant phase element) circuit in the mid-frequency region. Here, the R$_{ct}$ and CPE1 correspond to the charge transfer resistance and effective capacitance from double-layer, respectively. And, the R$_{i}$ and CPE2 correspond to resistance and capacitance of CEI film on the electrode surface, respectively. In the low-frequency region, W is the Warburg component, used to extract the diffusion of Na ions inside the active material, represented by a sloping line in Nyquist plot \cite{Zhang_JMCA_2018}. The values of R$_{ct}$ are obtained as 553.8, 377.8, 309.1, 229.3 and 363.2 $\ohm$ for the $x=$ 0, 0.01, 0.03, 0.05 and 0.07 cathodes, respectively. A lower R$_{ct}$ value for the $x=$ 0.05 demonstrate the formation of low energy ion-migration routes and construction of a conducting carbon coating lead to efficient ion and electron transfer. The apparent diffusion coefficient of the sodium ions across the interface can be obtained from the low-frequency region using below equations \cite{Cai_JPS_2023}:
\begin{equation}
D = \frac{R^{2}T^{2}}{2A^{2}\sigma^{2}n^{4}F^{4}C^{2}}
\end{equation}
\begin{equation}
Z_{re} = R_{s} + R_{ct} + \sigma\omega^{-0.5}
\end{equation}
where R is the gas constant, T is the temperature, A is the active surface area of the electrode, F is the Faraday constant, $n$ is the number of electrons participating in the redox reaction, C is the concentration of Na$^+$ in the electrode and W is the Warburg coefficient in the low-frequency region, related to the Z$_{re}$ inversely, in the above equation and is calculated from the linear fitting of Z$_{re}$ versus $\omega^{-0.5}$ plot, as depicted in Fig.~S5(d) of \cite{SI}. The $D$ values are found to be in the range of 10$^{-13}$--10$^{-14}$ cm$^{2}$ s$^{-1}$, which seems to be lower than obtained from CV and GITT, as discussed above. 

\begin{figure}
	\includegraphics[width=0.65\linewidth]{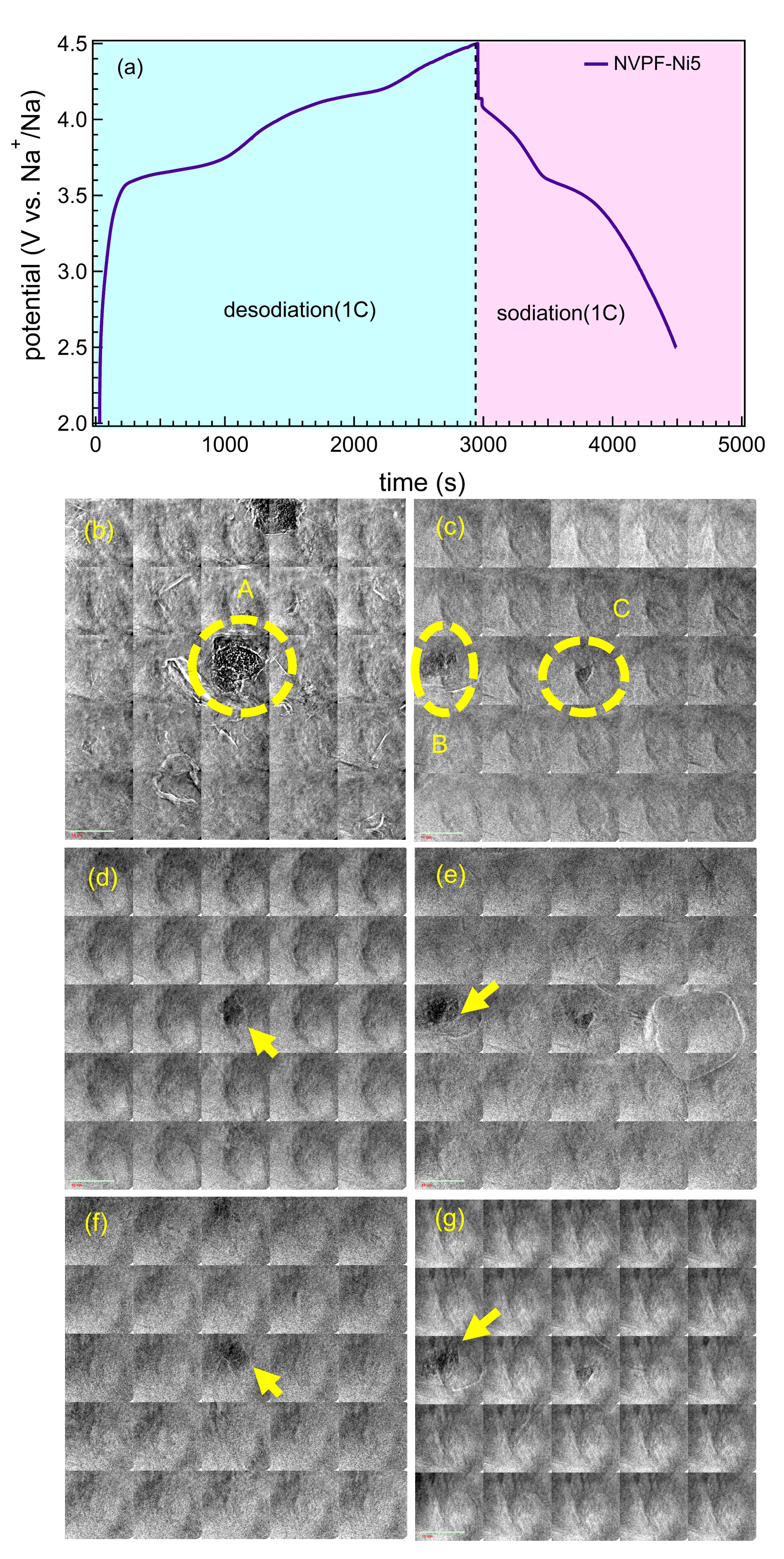}
	\caption{(a) The GCD profile of the {\it in-situ} TXM measurement at 1 C after cell formation for the $x=$ 0.05 cathode, (b, c) the mosaic patterns of the electrode at two regions showing particles (A, B, C); (d, e) the mosaic pattern during the charging and discharging (f, g) for the selected regions shown in panel (b, c), recorded after 60s during the charging and discharging process.}
	\label{TXM}
\end{figure} 
 
 The internal defects generated upon replacement of V$^{3+}$ ions by Ni$^{2+}$ increase the  formation energy, which accelerates the electron transfer and thus, increases conduction of Na$^{+}$ ions and electrons \cite{Cai_JPS_2023}. These results align very well with the polarization values, described in CV measurements. In addition, the electronic conductivities [see Fig.~S5(e, f)] of all the electrodes are calculated through the potentiostatic polarization method, and presented in \cite{SI}. The observed improved electrochemical performance with substitution of larger ionic radii Ni$^{2+}$ ions is consistent with the observed lattice expansion and broadening of the diffusion channels for the faster ionic transport during the sodiation/desodiation. The \textit{in-situ} carbon coated layer on the particles further enhances the surface conductivity and diffusivity for facile movement of Na$^{+}$ and electrons. The reduced particle size further shortens the diffusion distance for ion transport and enhances interactions at electrode-electrolyte interphase \cite{Wang_Nanoscale_2010}. These factors may result in observed excellent electrochemical performance, especially at high current densities for the $x=$ 0.05 optimized concentration of Ni doping. However, for the $x>$ 0.05, the electrochemical performance becomes sluggish with the reduced kinetics/diffusivity as visible from R$_{ct}$ values in EIS and enhanced polarization in GCD profiles \cite{Dee_AMI_2024}.  

\noindent 3.3 \textit{{\it Operando} XRD measurements: }\\
More intriguingly, to further elucidate the crystal structure evolution during the real-time electrochemical cycling, {\it operando} X-ray diffraction measurements are conducted for the $x=$ 0.05 electrode, which deliver insights of structural changes during the embedding/dis-embedding of Na$^{+}$ to/from the lattice framework. The half-cell is subjected to X-rays through a Be window during charging and discharging at a rate of 0.3 C. The {\it operando} 2D contour plot showing the evolution of diffraction peaks along with the charge/discharge profile and collected diffraction patterns are shown in Figs.~\ref{operando}(a, b). Interestingly, we observe the shift in the diffraction planes (220), (113) and (222) to the higher angles from OCV and (111) and (002) planes shift towards lower angles with the increase in the operating potential during the first cycle charging process. A total shift of 0.7\degree and 0.3\degree is observed for the (220) and (222) planes between OCV and the cut-off potential 4.5 V. Meanwhile, towards the end of charging at 4.5 V, the (113) plane merges with (220) appearing a single broad peak, which can be attributed to the the change in crystal structure and inter-planar spacing upon the Na$^{+}$ deintercalation \cite{Guo_JEC_2022}. Further, the calculated lattice parameters ($a, b, c$) and cell volume (V) are shown in Figs.~\ref{operando}(c, d), which clearly show gradual and reversible change due to oxidation/reduction of V$^{3+}$/V$^{4+}$ upon removal/insertion of Na$^{+}$ from/into the crystal lattice during charging/discharging process, and this behavior originates from the change of electrostatic force between F and O atoms in the [VO$_{4}$F$_{2}$] octahedron unit \cite{Deng_NanoEnergy_2021}. Importantly, this confirms the phase transition from Na$_{3}$V$_{1.95}$Ni$_{0.05}$(PO$_{4}$)$_{2}$F$_{3}$ to NaV$_{1.95}$Ni$_{0.05}$(PO$_{4}$)$_{2}$F$_{3}$ with extraction of Na$^{+}$ ions upon charging to 4.5 V as peaks (220) and (113) merge and generate one broad peak. On the other side, the diffraction planes (111) and (002) shift to the left during charging while the planes (202), (400) and (024) does not show any change in the 2$\theta$ values, as clearly seen in the Fig.~\ref{operando}(e--i). 

On the other hand, during the discharging process, the planes (220), (113) and (222) followed an opposite trend and began to shift towards the lower angles and once the discharging is complete, these planes revert back to their original positions of 27.8$\degree$, 28.5$\degree$ and 32.5$\degree$, respectively. Similarly, the diffraction planes (111) and (002) shift towards right on their respective positions. Notably, finite intensity of the (111) plane after cycling indicate the the structural stability of the crystal framework of the cathode during extraction of the Na ions during charging process. The results are consistent with the Bragg's law, the peak shifts correspond to the stepwise Na$^{+}$ ion insertion/deinsertion from the crystal lattice, yielding the phase changes \cite{Guo_JEC_2022, Zhou_Small_2024}. We performed 2-3 cycles of subsequent charging/discharging and found the shift reproducible as clearly seen in Fig.~\ref{operando}(f--i). Therefore, the reversible change of the structure confirms the high stability of the crystal lattice, which align well with the observed long cyclic stability and reversible electrochemical performance of the studied cathodes. 

\begin{figure*}
	\includegraphics[width=\linewidth]{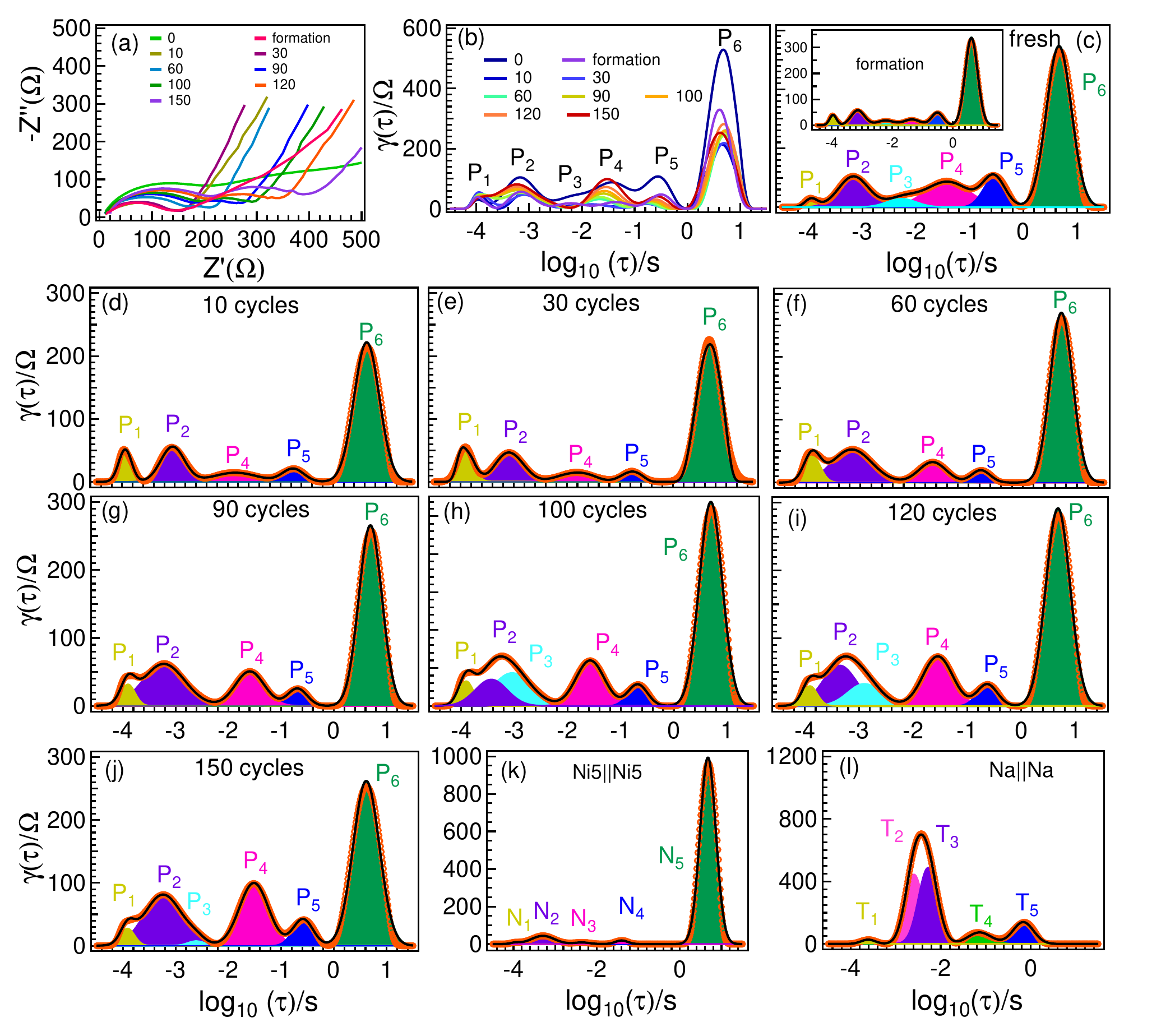}
	\caption{(a) The {\it in-situ} Nyquist plots recorded at 0.5 C for the $x=$ 0.05 cathode up to 150 cycles after the formation process; the DRT profiles with cycle life: (b) at selected cycles, (c) fresh cell (inset shows the formation process), (d) 10$^{th}$ cycle, (e) 30$^{th}$ cycle, (f) 60$^{th}$ cycle, (g) 90$^{th}$ cycle, (h) 100$^{th}$ cycle, (i) 120$^{th}$ cycle, (j) 150$^{th}$ cycle, (k) Ni5$\parallel$Ni5 and (l) Na$\parallel$Na symmetric cells.}
	\label{drt}
\end{figure*}

\noindent 3.4 \textit{{\it In-situ} synchrotron transmission X-ray microscopy: }\\
The {\it in-situ} synchrotron based transmission X-ray microscopy (TXM) measurements are conducted for the optimized $x=$ 0.05 (Ni5) concentration at 1 C after the cell formation to investigate the morphology evolution of the cathodes during the cycling. The monochromatic X-rays are directed onto the cathode material inside the coin-cell, where the steel covers are perforated and sealed through the Kapton tapes, and operation principles at the beamline are discussed in detail in \cite{Yang_FER_2018}. The radiation passing through the sample allows the direct visualization of the non-equilibrium/fast process in the active materials, without the relaxation periods through the {\it in-situ} TXM measurements. It applies the absorption and phase contrasts to unveil {\it in-situ} internal morphological variations of the heterogeneous sample, thereby grasping valuable insights into the battery performance \cite{Spence_Nanotechnology_2021, Bak_NPGAsia_2018}. Fig.~\ref{TXM}(a) presents the potential versus time plot of Ni5$\parallel$Na half cells during the desodiation/sodiation process and the corresponding 2D mosaic patterns highlighting the morphology evolution for the probed particles in the non-charged (b, c), charged (d, e) and discharged (f, g) states are shown in Fig.~\ref{TXM}. We observe that only few particles in the field of the view due to the electrode heterogeneity/configuration, which are represented by A, B and C (marked in yellow in panel b, c), in the mosaic patterns in Fig.~\ref{TXM}(b, c). The TXM image where the particle, A in its pristine state is shown in Fig.~S6(i) and for particle C, it is also shown in Fig.~S6 for the pristine (j), charged (k, l) and discharged (m) states in ref.~\cite{SI}. 

Interestingly, during the charge process, the particle C has undergone a slight displacement [see panel (k)] from its original position and as the desodiation continues, though the resolution is affected (see panel l) due to the increased side reactions at higher potentials, but the particle's morphology is still maintained with a slight change in lattice site, without any visible dissolution/surface cracks and volume changes. Therefore, it is concluded that Ni$^{2+}$ ion increases the volume expansion tolerance of the cathode with no contraction/expansion, and stabilizes the structure, which aligns well with the stable electrochemical performance. Furthermore, to glean insights into the dynamics of the particle morphology, the TXM images for a single particle are collected as a function of SOC and the snapshots at few potentials during the intecalation/deintercalation are shown in Fig.~S6(n, o) of \cite{SI} and video attached with the manuscript. However, there are still resolution issues due to the sample drifts, encountered during the non-trivial {\it in-situ} battery operations. The sample drifts due to the beam induced local heating and other side effects (zone plate resolution, low signal to noise ratio, depth of focus, contact issues etc.) \cite{Spence_Nanotechnology_2021}, which have significantly reduced the {\it in-situ} experiment efficiency. It can be clearly observed from the attached video that the particle is undergoing displacements from the original position, without any visible cracks or volume expansions, despite the prevalent stability issues affecting the image quality. In short, the {\it in-situ} TXM measurements provide a clue to correlate the microstructure with battery performance as a function of SOC and with further optimizations, some experimental issues can be resolved \cite{Weker_EES_2014}. 

\noindent 3.5 \textit{Distribution of Relaxation Times (DRT) Analysis: }\\ 
To investigate the evolution of the electrochemical reactions at different cycling lengths, the {\it in-situ} EIS measurements are performed while the coin cells of $x=$ 0.05 are subjected to continuous charge/discharge at 0.5 C rate. The Nyquist spectra are recorded for 0, 10, 30, 60, 90, 100, 120 and 150 cycles, as shown in Fig.~\ref{drt}(a). The charge transfer resistance at the electrode-electrolyte interface increases with the cycling, as depicted by two semicircles in the high-frequency regions. The straight line at the low frequencies depicts the bulk solid-state diffusion phenomenon of ions inside the active material, as discussed above. However, the major physical processes involved in SIBs originate from electronic conduction, ionic migration, physical contact, and activation and concentration polarizations, which cannot be studied through the impedance spectra due to the overlapping of the different processes in the same frequency range, owing to the complex microstructure of the electrodes \cite{Plank_JPS_2024}. Therefore, the study of distribution of relaxation times (DRT) is carried out for the better portrayal of each recorded EIS spectra in Fig.~\ref{drt}(b) to unveil the underlying electrochemical processes in the different time domains. These processes are represented by the peaks of different magnitude at particular time constants in DRT plot shown in Fig.~\ref{drt} (b). The profiles depict six peaks (P$_{1}$--P$_{6}$) where each peak has a characteristic time constant or frequency for different physical processes. The time constant ($\tau$) is a characteristic feature of the electrochemical process, and the area under the peak gives the value of the corresponding polarization involved \cite{Semerukhin_ECA_2024}. The peak, P$_{1}$ is denoted to the contact resistance between the particle--particle and particle to the current collector, while the peak, P$_{2}$ is attributed to the ionic transport processes across interface through the CEI and SEI conductive films. The peaks, P$_{3}$ and P$_{4}$ correspond to the resistance, originating from the secondary charge transfer reactions through SEI/CEI inside the electrochemical system. The peak, P$_{5}$ is attributed to the charge transfer process at the surface of the active materials and P$_{6}$ signifies the solid-state diffusion resistance of particles inside the cathode material. Notably, the peaks P$_{1}$ and P$_{2}$ lies in the high-frequency regions, while the peaks P$_{3}$ and P$_{4}$ fall in the mid-frequency regime and the peaks at P$_{5}$ and P$_{6}$ are covered in the low-frequency regions \cite{Lu_Joule_2022, Semerukhin_ECA_2024, Plank_JPS_2024, Manikandan_JPS_2017}.

In order to perform the DRT measurements, it is crucial to validate the EIS data provided by the electrochemical process and check for its linearity, causality, stability and finiteness \cite{Plank_JPS_2024}. The detailed information about the DRT theory and methods has been provided in our recent work \cite{Pati_JPS_2024}. So, the validity test is performed by evaluating the Kramers-Kronig (KK) relations, which relates the real and imaginary components and are verified using Lin-KK software \cite{Boukamp_JES_1995} and both the residuals are less than 2\%, as represented in Fig.~S7 of \cite{SI}. The evolution of the distribution function, represented by $\gamma$($\tau$) with the cycle life, is shown through the DRT profiles for the fresh and cycled cells (different cycles) in Figs.~\ref{drt}(c--j). The intensity of the distribution function is found to change with the cycling, thereby showing the variations in peak position and peak width. Herein, it is observed that the SEI formation in the fresh cell is not stable and incompatible for the absorption of the electrolyte, which stabilizes after the formation cycle, but initially increases the impedance of the cell, as depicted in Figs.~\ref{drt}(a, b) and Table~S8 of \cite{SI}. When the cell undergoes cycling at 0.5 C rate for 10 cycles after formation, the SEI becomes stable for the charge transfer and resistance decreases with negligible contribution from the secondary processes (P$_{3}$) as clearly depicted in Figs.~\ref{drt}(a, d). However, as the cycling continues, the polarization starts increasing again due to the thickening of SEI film and its intrinsic ageing mechanism \cite{Verma_ECA_2010}. As the cycle number exceeds 100, the effect of secondary ion transport also pronounces along with the internal and diffusion resistance of the Na$^{+}$ ions (see P$_{3}$ and P$_{4}$), which was initially present for the fresh cell and then got reduced after formation process, in Fig.~\ref{drt} (h). This phenomena occurs due to the loss of cyclable sodium during long cycling and further forms different decomposition species, which is associated with the growth and thickening of the SEI film. The variation of different polarizations for fresh, formation, 10, 30, 60, 90, 100, 120 and 150 cycles are depicted in Table~S8 of \cite{SI}. 

Further, to distinguish between anode and cathode contributions in the above calculated polarizations, the symmetric cells (Na$\parallel$Na and Ni5$\parallel$Ni5) are fabricated to study the physical processes and the deconvoluted DRT spectra are illustrated in Figs.~\ref{drt}(k, l) \cite{Wang_EnergyT_2022}. In the Fig.~\ref{drt}(k), the N$_{1}$ and N$_{2}$ peaks appear in the range of 10$^{-4}$--10$^{-2}$s, while N$_{3}$ and N$_{4}$ peaks lie in range of 10$^{-4}$--10$^{-2}$s, signifying the contact resistance and interfacial resistance are quite smaller values than the above configuration. This means that the additional polarization is compensated by the interfacial resistance from the Na metal/anode side. While, the strong diffusion peak N$_{6}$ observed between 0--10s, clearly matches well with the P$_{6}$ with enhanced polarization and confirms the bulk diffusion of the Na$^{+}$ ions inside the cathode material. Additionally, the DRT spectra of Na$\parallel$Na symmetric cell shows peaks T$_{1}$--T$_{5}$ in the range of 10$^{-4}$---1s and we observe that T$_{3}$, T$_{4}$ and T$_{5}$ match well with P$_{3}$, P$_{4}$ and P$_{5}$ peaks, thereby confirming the significant dominance of charge transfer resistance at anode interface during electrochemical cycling. Therefore, the study of relaxation times can help us identify the nature and magnitude of the physio-chemical processes and further tune the interfacial chemistry for developing efficient batteries. 

\noindent 3.6 \textit{Post--cycling {\it ex-situ} XRD and SEM measurements:}
Finally, the {\it ex-situ} XRD and SEM analysis of the electrode material are performed to study the structural and morphological evolution after 300 cycles. For this purpose, the cycled coin cells were dismantled, and cathodes were washed with dimethyl carbonate liquid solvent to remove the reactive species, followed by drying in vacuum oven at 60\degree C for 12 hrs inside the glovebox. The XRD patterns, depicted in Fig.~S8(a) of \cite{SI} indicate some traces of NVP phase in different concentrations along with some other secondary phases, as marked in Fig.~S8(a), which signifies the possible dissolution of the fluorine in the electrolyte upon cycling \cite{Dee_SCE_2021}. The dissolution is found to be small in the case of $x=$ 0.05 [see Fig.~S8(g)], while it is significant for other cathodes, which matches well the cyclic performance behavior with Ni concentration. Overall, the is no significant shifting of the peak positions illustrating minimal structural changes during battery cycling. Besides the dissolution, the intensity of few peaks in {\it ex-situ} XRD patterns is not pronounced due to the shielding from the interfacial layer formed on the electrode surface. The SEM images in Figs.~S8(b--g) of \cite{SI} suggest that the active materials aggregate during prolonged cycling with the thickening of SEI film and lattice volume changes, which results in the cracking of the electrode surface and eventually leads to capacity decay \cite{Dee_AMI_2024}. There are also a few glass fibers from the separator evident on the surface, which could not be entirely removed during the disassembly process. Also, the CEI interface becomes pronounced which further contributes to the generation of cracks for the pristine cathode. However, the optimized $x=$ 0.05 cathode show lesser degradation and crack formation, possibly due to the formation of stable passivation layer and promotes facile ionic transfer during long cycling. Herein, the dopant Ni$^{2+}$ prevents structural deformation during battery cycling by also acting as a pillar and maintains the structural stability, as clearly observed in the high resolution SEM image [Fig.~S8(g)] of \cite{SI}.

\section{\noindent ~Conclusions}

In summary, the Na$_{3}$V$_{2-x}$Ni$_{x}$(PO$_{4}$)$_{2}$F$_{3}$/C ($x=$ 0--0.07) series of cathodes were successfully designed and synthesized through the facile sol-gel route. The crystal structure, morphology, microstructure, and valence state have been confirmed through the X-ray diffraction, field emission scanning electron microscopy, transmission electron microscopy, X-ray photoelectron spectroscopy measurements. The optimally doped $x=$ 0.05 cathode exhibit superior specific capacity of 119 and 100 mAh g$^{-1}$ at 0.1 C and 10 C rates, respectively. Interestingly, it also showed excellent capacity retention of 78\% after 2000 cycles at higher 10 C rate with nearly 100\% Coulombic efficiency. The apparent diffusion coefficient values, obtained from CV and GITT, lie in the range of 10$^{-9}$--10$^{-10}$ cm$^{2}$ s$^{-1}$. Notably, the inclusion of Ni$^{2+}$ dopants in NVPF cathode has effectively reduced the charge transfer resistance, electrochemical polarization and increased the diffusion pathways. More importantly, the structural evolution during the cycling was investigated through the lab-based {\it operando} X-ray diffraction measurements, which confirm their reversible crystal structure. A short study of the microstructure evolution through synchrotron {\it in-situ} transmission X-ray microscopy proved the structure stability, especially at high potentials. The study for the distribution of relaxation time (DRT) with the evolution of cycle life helped to understand the capacity fading and obtain more precise polarization values from physical processes in the applied frequency range. Therefore, this study provides a benchmark for the development of robust fluorophosphate cathodes in next generation high-performance sodium-ion batteries.  

\section{\noindent ~Conflict of interest }
The authors declare that there are no conflicts of interest associated with this article. All data presented are original contributions and have not been concurrently submitted for publication elsewhere.

\section{\noindent ~Acknowledgements}
Simranjot K. Sapra would like to thank IIT Delhi and NYCU Taiwan for the fellowship. We thank the National Science and Technology Council (NSTC) of Taiwan for providing the instrumentation facilities for the characterization of the developed materials. SKS thanks Physics department, IIT Delhi for XRD measurements. We thank NSRRC Taiwan for providing beamtime for synchrotron XRD and TXM measurements. SKS thanks Prof. Diah Augstina Puspitasari for the fruitful discussions. SKS also thanks Luke, Rahmandhika Firdauzha Hary Hernandha, Nindita, Ryan and Bharath Umesh for the help during the experiments. RSD acknowledges the Department of Science and Technology, Government of India, for financial support through ``DST--IIT Delhi Energy Storage Platform on Batteries" (project no. DST/TMD/MECSP/2K17/07) and SERB--DST through a core research grant (file no.: CRG/2020/003436). RSD and JKC thank NYCU and IIT Delhi for the support through MFIRP project (MI02683G).

\end{document}